# Accessible Computer Science for K-12 Students with Hearing Impairments

Meenakshi Das[1] Daniela Marghitu[1] Fatemeh Jamshidi[1]

Mahender Mandala[2] and Ayanna Howard[2]

[1]Auburn University, Auburn AL 36830, USA
[2]Georgia Institute of Technology, Atlanta GA 30332, USA

An inclusive science, technology, engineering and mathematics (STEM) workforce is needed to maintain America's leadership in the scientific enterprise. Increasing the participation of underrepresented groups in STEM, including persons with disabilities, requires national attention to fully engage the nation's citizens in transforming its STEM enterprise. To address this need, a number of initiatives, such as AccessCSforALL, Bootstrap, and CSforAll, are making efforts to make Computer Science inclusive to the 7.4 million K-12 students with disabilities in the U.S. Of special interest to our project are those K-12 students with hearing impairments. American Sign Language (ASL) is the primary means of communication for an estimated 500,000 people in the United States, yet there are limited online resources providing Computer Science instruction in ASL. This paper introduces a new project designed to support **Deaf/Hard of Hearing (D/HH) K-12 students and sign interpreters in acquiring knowledge of complex Computer Science concepts. We discuss the motivation for the project and an early design of the accessible block-based Computer Science curriculum to engage D/HH students in hands-on computing education.**

## 1. Introduction

Accessibility of STEM topics is a topic that has actively engaged the Human-Computer Interaction (HCI) research community, which aims to ensure that STEM applications and services can be used on an equal basis by users with disabilities and underrepresented groups [1]. Initiatives such as AccessCSforALL [2], Bootstrap [3] and CSforAll [4] are active projects focused on engaging students with disabilities in Computer Science (CS) educational opportunities. For the deaf and hard of hearing (D/HH) community, initiatives such as ASLCORE [5] from the National Technical Institute for the Deaf (NTID) at the Rochester Institute of Technology (RIT) and ASLClear [6] are working on creating ASL signs for STEM disciplines.

American Sign Language (ASL) is a language very distinct from English and contains "all the fundamental features of language, with its own rules for pronunciation, word formation, and word order" [7]. A study by Heunerfauth and Hanson revealed that Deaf individuals often acquire sign language as their first language, and they are most fluent and comfortable in it [8]. They stated that sign language interfaces are a necessity for such D/HH individuals. Websites such as YouTube provide closed-captioning features for videos, but the captions are often inaccurate, and the reading level of this text is sometimes too complex and difficult for deaf individuals. Half of deaf students pass from secondary school with a fourth-grade reading level or less [9]. Another study stated that the frequently reported low literacy levels among students with severe hearing disability were partly due to the discrepancy between their "incomplete spoken language system and the demands of reading a speech-based system" [10]. This illustrates the need of providing coding instruction in ASL. D/HH students are at a high risk of unemployment or chronic underemployment at a rate of 70% or more nationally [11]. The number of tech job openings in the country is growing exponentially and hence, by encouraging D/HH students to learn CS concepts and tackle programming challenges from the middle school level using accessible coding instruction, we can help to lower the unemployment rate.

To address this need, in this paper, we discuss a project designed to 1) bridge the gap between K-12 D/HH students and an active block-based coding learning environment, 2) expose students to CS topics within a supportive and fun environment, including robots, and 3) provide age-appropriate ASL video resources for K-12 students while encouraging independent learning and creative problem solving.

## 2. Related Work

Despite the fact that human signing videos are expensive and hence difficult to update, there is a need to produce sign language content explaining core computer programming concepts which don't require frequent updating. For example, in the CS field, ASLCORE has produced signs and definitions for vocabulary and concepts such as "Recursion", "Debugger", "Linked List" and "Variable". In this way, CS jargon is being made accessible to older students who are D/HH, who have to otherwise rely on fingerspelling while communicating with this vocabulary. Apple has also recently released seven videos [12] explaining computer programming concepts in ASL. On the other hand, Drag and Drop coding applications such as MIT's Scratch [13] are popularly used to teach younger students to code. Our project aims to make CS principles, using Block-based coding interfaces, a more inclusive experience for D/HH students since many students' first experience with programming is through such coding interfaces.

A number of ambitious projects (e.g. automatic sign language recognition, animated 3D humanlike characters who perform sign language, and reading assistance technologies) are being developed to make web content more accessible to D/HH students. Although substantial work [8] is being done in creating this animated sign language using avatars, which automatically converts a body of text into avatars signing them, the majority of the work is still in the research phase. Our project does not aim to create sign language videos for every content related to CS on the web; automated virtual human sign technologies, when fully developed, are a better fit for this purpose. Instead, our objective is to create engaging human signing videos for core computing science concepts and common data structures in the context of a block-based programming environment. There are two primary reasons for doing this:

> **1. To increase *pre-lingual* deaf students' excitement about CS.**
> According to the Communication Services for the Deaf, "98% of deaf people do not receive education in sign language" [14]. Lisa Harrod, a former sign language interpreter, mentions that captioning is simply the written form of spoken words which can be very difficult for D/HH who struggle with English as a second language [15]. Pre-lingual individuals (born with deafness or became deaf very young) who learned ASL as their first language will more likely be interested to learn coding through ASL, rather than just reading paragraphs of text or captioned videos explaining computer programming concepts. Therefore, this project tries to provide a more promising and engaging learning environment for these students that possibly their second language could not. In addition, we hypothesize that a student's sense of inclusion will increase with the knowledge that computer programming is being taught in a language that addresses their needs.
>
> **2. To improve *post-lingual* deaf students comfort level with common CS ASL signs.**
> For post-lingual students who learned ASL later in life and are not very comfortable with it, web applications should provide captioning and/or transcripts for every ASL video, so that they do not have to solely depend on ASL. However, our plan is to familiarize post-lingual deaf students with the basic ASL signs for programming concepts, so they may use it to communicate with other deaf students, in addition to providing captioning in videos.

## 3. Project Conceptual Details

This project is part of a larger research initiative which aims to develop an accessible block-based web application for middle school students with disabilities. The app aims to provide accessibility features for students who are either blind/visually impaired, D/HH, have physical, and/or cognitive disabilities. This paper particularly focuses on the accessibility considerations for the D/HH audience, which is further divided into two parts. The first part discusses the development considerations for videos explaining core CS concepts in ASL, which will be embedded as part of the curriculum for the accessible web application project. The signs for STEM terminologies will be based on ASLCORE to ensure technical accuracy. Our ASL videos explain the following programming concepts with real-life application examples: **a.** Introduction to CS & Programming **b.** Pseudocode & Algorithms **c.** Commands **d.** Conditional Statements & Event Driven Development. **e.** Loops **f.** Variables **g.** Functions **h**. Debugging. **i.** Object-Oriented Programming. These concepts are inspired by materials from CS Teachers Association's K-12 CS Standards, and the CS4ALL curriculum developed for an inclusive robot camp at Auburn University [16].

It is proven that D/HH individuals are visual learners [17], hence every video will demonstrate application of the particular programming concept through a block-based coding interface. Videos for other CS concepts which explain the social aspect of computing, such as cybersecurity, bias in computing, accessibility, and careers in computing, are also being filmed. We have received a SIGCSE special projects grant [18] to produce these videos. Hence, we have partnered with the non-profit organization Deaf Kids Code [11] to recruit D/HH STEM teachers and successful deaf technologists to star in these videos. Another key reason to specifically produce these computing videos in ASL is to avoid the denotation and connotation confusion of English words [19], which D/HH individuals sometimes experience. For example, the word 'call' in English has the most common meaning of contacting someone or giving someone a name. However, in the CS context, it means to execute a subroutine or a function. Young D/HH kids, especially those in middle school, may not have the contextual knowledge to compensate for these informational gaps [19], hence explaining and clarifying these contextual differences through ASL videos would be extremely helpful for D/HH middle-schoolers. Additionally, research shows that the majority of teachers for the D/HH are hearing. 2,575 K-12 D/HH teachers were surveyed and it was found that less than 22% were deaf [20]. Only 38% of D/HH teachers believed that "their expressive sign skills were on par with their expressive spoken English" [20]. This suggests that only a few D/HH K-12 individuals have access to teachers who communicate with them using ASL, let alone using appropriate signs for STEM terminologies. Hence, our videos can be used by STEM teachers in their classrooms to introduce students to CS concepts.

The second part of this project is developing built-in accessibility features for D/HH audience for the Block-based coding web application. The web application consists of some programming tutorials similar to Activity Guides [21] in MIT's Scratch. Additionally, it consists of an extensive robot platform where students can use programming blocks to control a robot. The benefits of learning CS through robotics are well researched [22]. In this sense, several robotic environments are being proposed to facilitate programming teaching, such as Lego Mindstorms [23] and NaoBlocks [24]. The Finch robot designed to align with the learning goals and concepts taught in introductory CS courses is extremely popular [25]. Our robot, which is designed at the Georgia Institute of Technology, is cost-effective and visually appealing to middle schoolers. It also has the capability to produce sounds which can serve as feedback for blind/visually impaired individuals. The robot has a wireless connection and commands to control it, via blocks, can be sent through Bluetooth from the web application.

Figure 1 is a picture of the top-view of our robot.

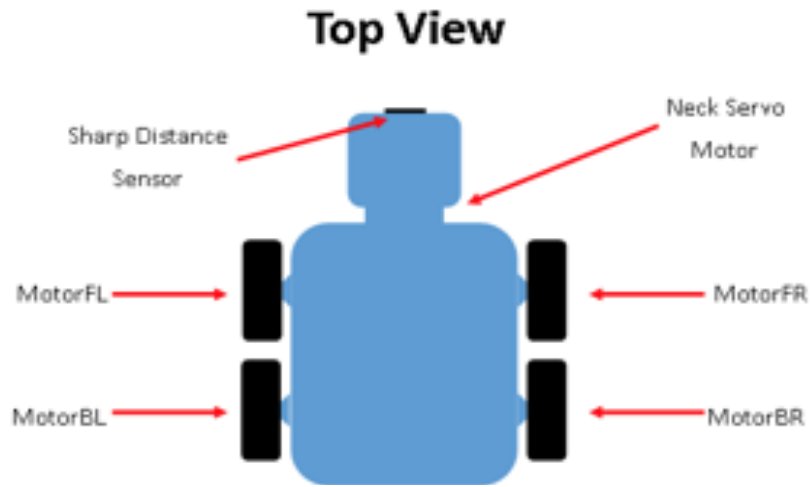

Figure 1: Top View of the robot showing the Sharp Distance Sensor, Neck Servo Motor, Front-left Motor, Back-left Motor, Front-right Motor & Back-right Motor.

Figure 2 is a picture of the bottom-view of our robot.

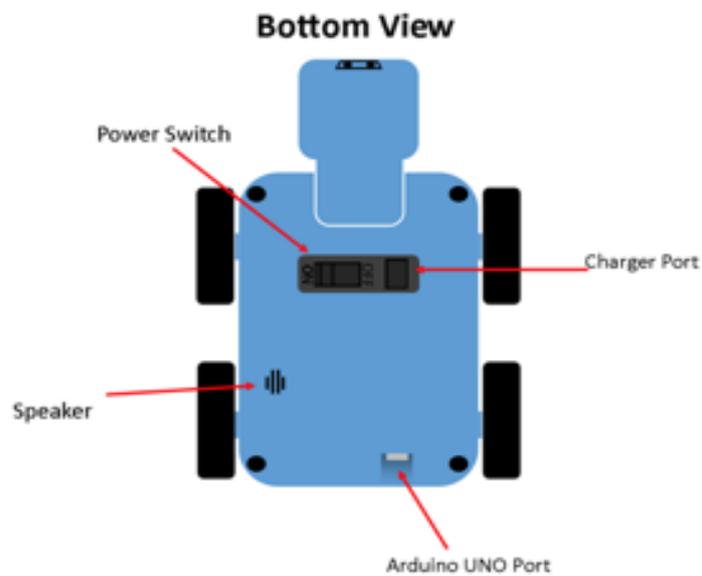

Figure 2: Bottom View of the robot showing the Power Switch, Speaker, Charger Port and the Arduino UNO Port.

The guidelines for the heuristic evaluation for Deaf Web User Experience (HE4DWUX) were used to develop a product requirements list for D/HH accessible web application [26]. This heuristic evaluation is based on the well-known usability heuristics developed by Jakob Nielsen [27]. The HE4DWUX consists of similar heuristics as Nielson's and provides additional context with regards to accessibility for the D/HH.

We discuss the following accessibility features of our app with regards to seven relevant heuristics described in the HE4DWUX.

1. **Sign language video content—The website should support captioned video content, alternative text and make use of signers in videos.**

The ASL videos will be embedded in the web application. Figure 3 is a prototype of this feature in action. The ASL video for this particular tutorial would explain the concept of commands such as 'turn left', 'turn right', and 'move forward'.

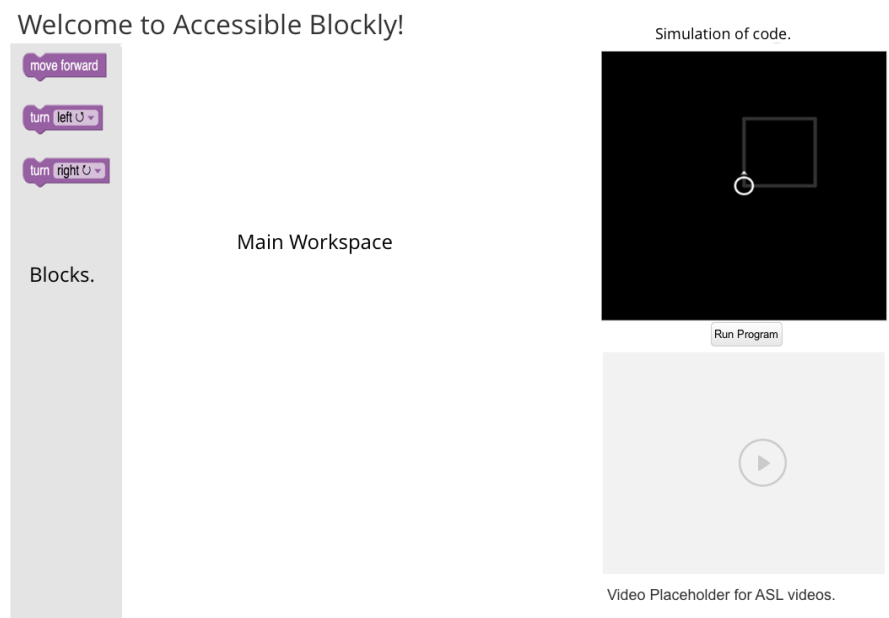

Figure 3. Under-development block coding interface showing Placeholder for ASL videos. Block and simulation images taken from Blockly Games.

2. **Visibility of status and actions—The website should provide users who are deaf with timely visual-based feedback on actions.**

Good visual feedback is very important to ensure web accessibility for D/HH. For example, when a user clicks on the run button to execute their code, it is necessary to provide clear visual cues such as text or graphics on screen, which can convey that the program has been run, as opposed to just audio cues. Some other examples of visual-based feedback in consideration are ensuring that content changes are indicated near the area of interest, using alternatives to static and textual forms of feedback, and text with icons for better navigation [26].

3. **Match between website and the world of the Deaf—The website should be aligned with language and concepts from Deaf culture.**

Our signs for programming concepts are heavily based on the work done by ASLCORE [5] since they have done extensive research in creating signs for technical programming terms such as "IF", "WHILE" and "FOR". Our partnership with Deaf Kids code allows us to understand the needs and expectations of D/HH students. ASLCORE recruits a 4-person all Deaf translation team who have strong and intuitive knowledge of ASL [28]. Similarly, only ASL experts will participate in the making of our videos to ensure authenticity and correctness. The web application also provides a feedback form so that users may let us know of any potential issues in the videos.

4. **User control and freedom—The website should make users who are deaf feel in control of their experience.**

Users are able to control the speed of ASL videos and increase the size of captions. ASL signs/videos for every drag and droppable 'block' like "IF-ELSE" or "WHILE", are displayed as *tooltips* for a block. Tooltip is a common graphical user interface element. It is used jointly with a cursor, commonly a pointer [29]. A research study conducted with 15 deaf users found that sign language and picture tooltips were very positively rated as opposed to only text-based tooltips or a human speaking-based tooltip (used for lip-reading) [30]. In our web application, once a user hovers their cursor over a block, it will play the sign for the block in ASL. Figure 4 is a prototype of this feature in action.

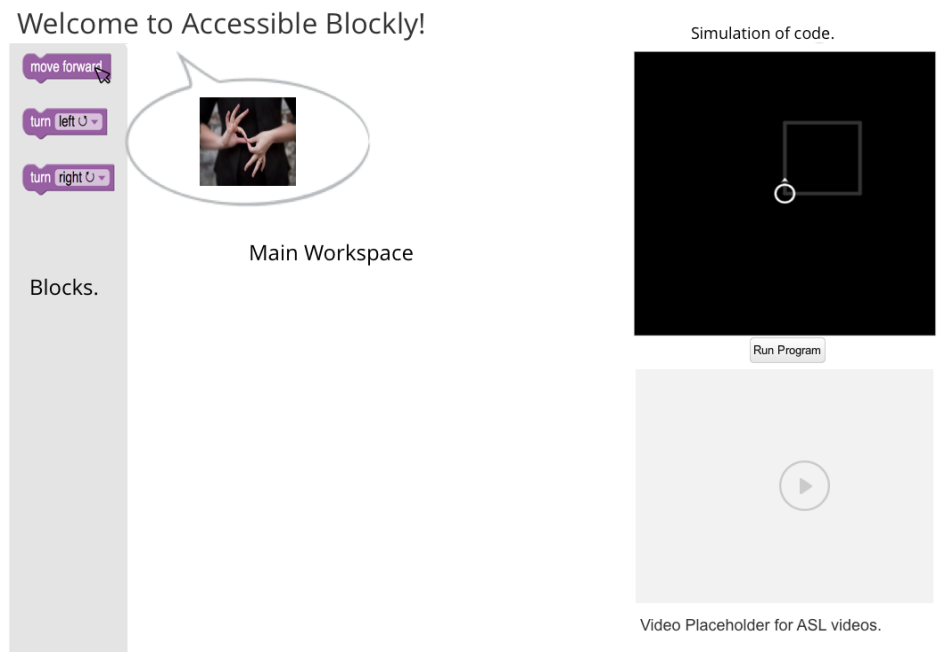

Figure 4. Under-development block coding interface displaying tooltips in ASL as a user hover over a block. Block and simulation images taken from Blockly Games. Handshape image taken from Calliope Interpreters.

However, if the user would prefer text as tooltips instead of ASL, they have the option to change to text view as well. This ensures users have control on the features of the web application. Figure 5 is a prototype of this feature in action.

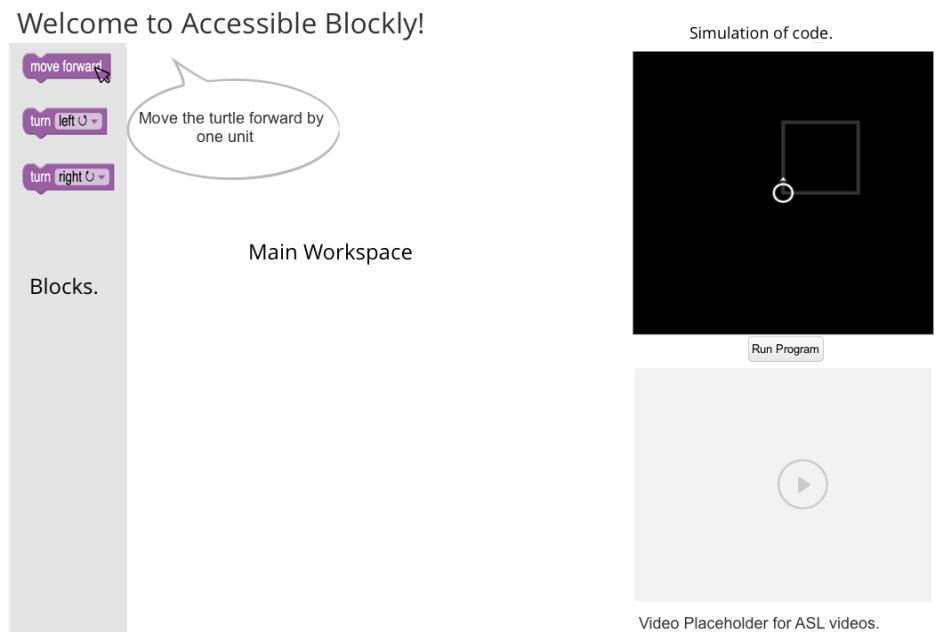

Figure 5. Under-development block coding interface displaying tooltips in English as opposed to in ASL, if user prefers English instead. Block and simulation images taken from Blockly Games.

5. **Captions—The website should describe dialogue and sound effects (audio content).**

Lisa Harrod also mentions, "Captioning is perfect for the post-lingual deaf or hard-of hearing audience; it presents content in an accessible format, in the primary language of the user" [15]. Every ASL video in our web application includes captions to account for post-lingual D/HH individuals who prefer captions.

6. **Aesthetic and minimalist design—The website should contain the most essential elements and any non-relevant or rarely needed content that will be competing with other units of content for visibility and attention must be removed.**

Our application aims for web content accessibility conformance (e.g. using sufficient color contrast, allowing zooming, minimizing distracting and cluttering content, and using proper headings). Hence, visual noise is eliminated to reduce memory overload. This will also be beneficial to individuals with cognitive disabilities.

7. **Personal skills—The website should support the skills and background knowledge of users who are deaf and not replace these.**

We incorporate multimodal learning in our app which is beneficial to D/HH individuals. Multimodal approaches "involve the use of multimedia along with the content text and encourage the use of interactive literacy practices between the instructor and the students". Lacina and Mathews researched how multimodal learning tools affect the learning process, and "found that it provided a pleasant experience, captured students' attention, increased their motivation of reading, and enhanced their literacy skills" [31]. Deaf explore the world mostly by the sense of sight and vision. "Graphics, pictures, videos, and the different forms of multimedia can facilitate learning of complex theories and generalizations, which are usually challenging for deaf students that do not know how to use sign language" [32]. Involving multiple representations of the educational materials can help in decreasing the loss of the received data. "These multimodal representations for deaf students can combine text and sign language, or visual displays (e.g., pictures or videos) along with sign language subtitles" [32]. There is a diversity of learning styles among normal students, they could be verbal, visual or a different type of learners. Therefore, it's crucial to include all possible learning styles for deaf students' e-learning techniques [33, 34].

## 4. Future Work

We are currently in ongoing conversations with ASL subject matter experts from Deaf Kids Code and working on our video scripts and storyboarding. We plan to start producing these videos in March 2020 and finish them by mid-May 2020. After the videos are ready, we will conduct a study with a middle-school deaf population to extensively examine the impact of ASL videos in learning CS. We understand not every deaf student has access to an ASL teacher familiar with CS. Therefore, we will release our videos online so any deaf student can access it. In addition, we plan to partner with more schools for deaf in the United States to share our work and conduct similar workshops. In addition to this, work is being done on the above-mentioned block-based web application. We are currently surveying teachers of students with blind/visually impairments, D/HH, physical and cognitive disabilities to establish and validate the appropriate accessibility requirements for the various disabilities.

## Acknowledgment

This material is based upon work supported by the National Science Foundation under Grant No. 1849101. The views expressed in this article are of the authors and do not necessarily represent the views of National Science Foundation, Auburn University, or Georgia Institute of Technology.